\documentclass[useAMS,usenatbib]{mn2e}

\usepackage{graphicx}
\usepackage{ulem}
\usepackage{float}
\usepackage{psfig}

\title[Intergalactic magnetic fields in Stephan's 
Quintet]{Intergalactic magnetic fields in Stephan's 
Quintet}
\author[B. Nikiel--Wroczy\'nski et al.]
{B. Nikiel-Wroczy\'nski$^{1}$\thanks{E-mail:iwan@oa.uj.edu.pl},
M. Soida$^{1}$,
M. Urbanik$^{1}$,
R. Beck$^{2}$, and
D.~J. Bomans$^{3,4}$\\
$^{1}$Astronomical Observatory, Jagiellonian
University, ul. Orla 171, Krak\'ow PL 30-244, Poland\\
$^{2}$Max-Planck-Institut f\"ur Radioastronomie, Auf dem
H\"ugel 69, 53121 Bonn, Germany\\
$^{3}$Astronomisches Institut, Ruhr-Universit\"at Bochum,
Universit\"atsstrasse 150, 44801 Bochum, Germany\\
$^{4}$Ruhr-Universit\"at Bochum Research Department 
``Plasmas with Complex Interactions''
}

\begin{document}

\date{Accepted 2013 July 4.  Received 2013 July 2; in original form 2013 March 13}

\pagerange{\pageref{firstpage}--\pageref{lastpage}} \pubyear{xxxx}

\maketitle

\label{firstpage}

\begin{abstract}
We present results of the VLA radio continuum total power and polarised intensity observations of Stephan's
Quintet at 1.43 and 4.86\,GHz along with complementary 4.85 and 8.35\,GHz Effelsberg observations. Our study
shows a large envelope of radio emission encompassing all the member galaxies and hence a large volume of intergalactic
matter. Infall of the galaxy NGC\,7318B produces a ridge of intergalactic, polarised emission, for which
the magnetic field strength was estimated as $11.0 \pm 2.2\,\mu$G, with an ordered component of $2.6 \pm 0.8\,\mu$G. The
energy density of the field within the ridge area is of the same order as  estimates of the thermal component, 
implying a significant role of the magnetic field
in the dynamics of the intergalactic matter. We also report that the tidal dwarf galaxy candidate SQ-B possesses a
strong and highly anisotropic magnetic field with the total strength being equal to $6.5 \pm 1.9\,\mu$G and an ordered component
reaching $3.5 \pm 1.2\,\mu$G, which is comparable to that found in normal-sized galaxies.
\end{abstract}

\maketitle

\section{Introduction}
\label{intro}

Intergalactic magnetic fields are among the least studied phenomena related to galaxy
groups. Only a limited sample of such objects has yet been studied at radio 
wavelengths. All these studies (e.g. \citealt{xuold}, \citealt{giaci}) focused on the
total power (TP) emission, not taking the polarised intensity (PI) into account.

Detection of intergalactic polarised emission is
an important issue.
Polarisation is caused by magnetic fields showing some degree of
ordering. This may mean either genuinely unidirectional fields (called
"regular", no reversals of its lines) or by squeezed/stretched random fields,
having a preferred direction of fluctuations but frequent reversals. The
latter  are called "anisotropic" fields. The existence of a unidirectional field suggests that it
originated in galaxies
hosting large-scale dynamos. Such a magnetic field is recognised by non-zero Faraday rotation
measures (RM), while a lack of measurable RM indicates the generation
of a twisted magnetic field, possibly compressed by intergalactic shocks \citep{reversal}.
The discrimination is possible by determining the RM via
multifrequency polarisation observations or by using the RM synthesis method \citep{rmsynth}.
Moreover, the polarised emission provides an extremely sensitive tool
to reveal possible gas compression and shearing flows that cause
the magnetic field to be aligned along the compression front
and/or perpendicular to the velocity gradients (\citealt{urb}, \citealt{mawez}).

Stephan's Quintet (located approximately 85\,Mpc from the Milky Way,
\citealt{distance}) is one of several galaxy groups that possibly host intergalactic 
magnetic fields, with magnificent dust tails emerging from NGC\,7319. Additionally, it contains a radio ridge
produced by the interactions with infalling NGC\,7318B galaxy \citep{xuold}, as well as a large-scale
{\rm H}{\sc i} tail overlapping the interloper galaxy NGC\,7320 \citep*{hi}. 

Since its discovery in 1877, Stephan's Quintet, named after
its discoverer \`Edouard Jean-Marie Stephan, has been subject to extensive,
multiwavelength studies, much more detailed than for any other compact group.
Denoted HCG\,92 \citep{hickson}, the group is known to exhibit numerous
interaction-related phenomena, such as changes in galaxies' morphology,
starburst activity (\citealt{xuold}), gas outflows visible in various regimes
(\citealt{hi}, \citealt{xuold}, \citealt{natale}, \citealt{guillard}), and possible
shock compression (\citealt{jabko}, \citealt{osull}). The group
is also clearly visible in radio continuum \citep{hi} and a
careful study of the New VLA Sky Survey (NVSS) data shows some hints for polarised
emission at 1.4\,GHz \citep{nvss}.

Several radio continuum studies of the Quintet and its member galaxies have 
been performed since 1972 (\citealt{arp}, \citealt{allen}). 
The most recent ones are the high-resolution study of the active galactic nuclei
in NGC\,7319 \citep{aoki} and that of the extended emission at 1.43 and 
4.86\,GHz \citep{xuold}. The authors of the latter study presented images 
of the TP emission, made using the Very Large Array (VLA) interferometer in its 
B and C configurations. These images show a large-scale radio emitting ridge between NGC\,7318B
and NGC\,7319 at both wavelengths, coincident with a UV emitting region and
X-ray features (presented by \citealt{trinchold, trinchnew} and
\citealt{osull}). 
Their high--resolution configurations of the VLA caused a substantial flux 
loss, and most of the extended emission was not detected. Moreover, the weak polarised 
emission (marginally visible in the NVSS map) remained undetected.

The distribution of the magnetic field (together with the X-ray morphology) of the 
Quintet has recently been simulated by \citet{simula}. Both models adapted 
by these authors (one from \citealt*{renaud}, the other from \citealt{hwang})
suggest a collisional origin of the shock region and a significant magnetic field 
between NGC\,7319 and NGC\,7318B.

In this paper we present observations of Stephan's Quintet made using the
VLA D-array, sensitive to the extended emission with a special attention paid to
polarisation. The observations were performed at two different frequencies: 1.43
and 4.86\,GHz. Additionally, we made single-dish observations
using the \mbox{100m} Effelsberg radiotelescope at 4.85 and 8.35\,GHz.

\section{Observations and data reduction}
\label{observs}

\subsection{Interferometric observations}

The 4.86\,GHz data were obtained in August 2008 using the Very Large Array (VLA) of the National
Radio Astronomy Observatory (NRAO)\footnote{NRAO is a facility of National Science
Foundation operated under cooperative agreement by Associated Universities, Inc.} 
in the D-array configuration. The total on-source time (TOS) was 21.5 hours. The 1.43\,GHz
data were also acquired using the D-array, with TOS of 4 hours; moreover, we have been
granted 3.5 hours in dynamic time allocation mode (CD- and D-array),  obtained between
February and April 2007. In both cases the bandwidth was 2$\times$50 MHz,
centred at 4835 and 4885\,MHz in C-band, and 1385 and 1465 in L-band.
                                 
The data were reduced using the Astronomical Image Processing System (\textsc{AIPS}) and 
calibrated using 3C\,48 at 1.43 and 3C\,286 at 4.86\,GHz as the flux and polarisation
position angle calibrators. The nearby point source 2236+284 was used as a phase and 
instrumental polarisation calibrator. For the 4.86\,GHz data, we made a set of Stokes
I, Q, and U maps using Briggs weighting (robust parameter\,=\,3), yielding a beamsize of 
$13.65\times 12.28$\,arcsec. These maps were later convolved to a circular beam of 
20\,arcsec. We also produced a uniformly weighted map of Stokes I channel, with a beam
of 6.8\,arcsec. The lower resolution set was used to produce distributions of diffuse 
TP and PI emission, while the uniformly weighted map shows details 
of the TP emission. At 1.43\,GHz maps in all Stokes parameters were convolved
to a common beam of 42\,arcsec. Finally, the U and Q maps at both frequencies
were combined to yield the distributions of polarised intensity (PI) and polarisation angle. 

\subsection{Single dish observations}

Single-dish mapping of the Quintet was done using the \mbox{100m} radiotelescope at 
Effelsberg\footnote{Based on observations with the \mbox{100m} telescope of the 
MPIfR (Max-Planck-Institut f\"ur Radioastronomie) at Effelsberg}. Observations 
were performed at 8.35\,GHz, using a single-beam receiver installed 
in the secondary focus of the telescope. The bandwidth was 1.1\,GHz and the final 
resolution (after some convolution) is 85\,arcsec.  In order to produce the final
map, 27 coverages were obtained, each of size $16\times16$\,arcmin, scanned
alternatively along R.A. and Dec. The scanning velocity was 30\,arcsec\,$\mbox{s}^{-1}$ and the
grid spacing was 30\,arcsec. All coverages were combined in the Fourier domain to reduce 
the scanning effects \citep{bweav}, for the Stokes parameters 
I, Q, and U separately. Again, we obtained maps of the polarised intensity and polarisation angle
from our combined U and Q data. The flux density scale was established using the source 3C\,286, 
according to the flux values given by \citet{baars}.
Additional $4.85$\,GHz mapping has also been
performed in order to provide the zero-spacing information that is missing in the
interferometric data. These complementary single-dish observations yielded
no larger integrated total flux, indicating that there were no flux losses in the interferometric
data, so that no merging was performed.

As the uncertainties of the flux values based on the r.m.s. noise levels
have turned out to be small compared to the
uncertainties of the calibration, we assume a 5 per cent error for
each integrated flux value for the radio maps.
The noise levels obtained for all radio maps are presented in Table \ref{rms}.

\begin{table}
\caption{\label{rms}Noise levels ($\sigma$) obtained in the final total power and
polarised intensity maps}
\begin{center}
\begin{tabular}{rrrrrrrl}
\hline
\hline
Freq. & $\sigma$(TP)&\multicolumn{2}{c}{$\sigma$(Stokes}&$\sigma$(PI)&beam&telescope/ \\
&&\multicolumn{2}{c}{ Q and U)} & &size&config.\\
~[GHz]&[${\mu\mathrm{Jy}\over\mathrm{beam}}$]&\multicolumn{2}{c}{[${\mu\mathrm{Jy}\over\mathrm{beam}}$]}&[${\mu\mathrm{Jy}\over\mathrm{beam}}$]&[arcsec]&\\
\hline
8.35 & 100 & 48 &54& 71 & 85   & Effelsberg\\
4.86 & 6  &  6 & 6 & 6  & 20   & D-array\\
1.43 & 110  & 22 & 24 & 32 & 42   & CD\&D-array\\
\hline
\end{tabular}
\end{center}
\end{table}

\subsection{The X-ray map}

To further characterise the magnetic field in the ridge (which is considered to be formed by
shock compression, \citealt{osull}), a map of emission in the
X-ray regime has been created using archival data from \textit{CHANDRA} 
(Program no. 7924, PI: \citealt{vrtilek}). The data were taken from the Project Archive
and then reprocessed using {\sc CIAO} software
ver. 4.3. The data were collected in the VFAINT mode, applicable for tracing 
the weak extended emission. The set was cleaned for bad events; the
total exposure time is 93.2\,ks. We have created an image in the soft X-ray regime 
(0.4--3\,keV), later smoothed using the {\it csmooth} tool.

\section{Results}
\label{result}
In this paper we use the term ``apparent polarisation B--vectors'', defined 
as the observed polarisation E--vectors direction rotated by $90\degr$, uncorrected for the Faraday
rotation, except for the maps in Sect.~\ref{discus}.

\subsection{Total power emission at 4.86\,GHz}
\label{sec6cm}
\begin{figure}
\resizebox{\hsize}{!}{\includegraphics{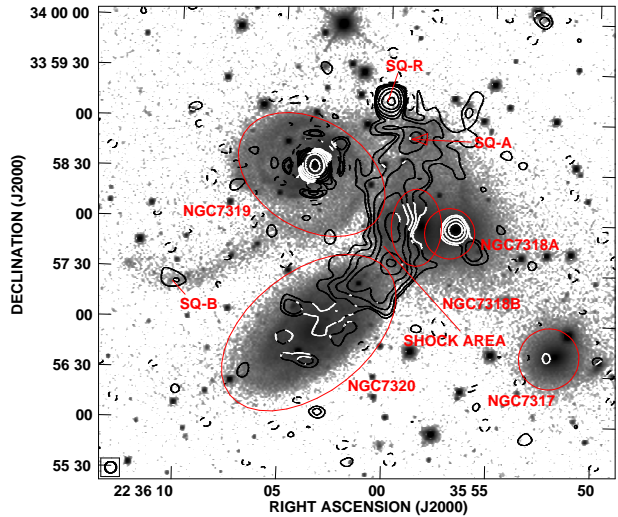}}
\caption{
Uniformly weighted VLA D-array map of the total power emission of Stephan's Quintet (SQ) at 4.86\,GHz overlaid upon an SDSS-R image.  
The contour levels are $-3$ (dashed), $3,5,10,20,50,100,200,500\times 10\,\mu \mathrm{Jy/beam}$
(r.m.s. noise level). The angular resolution is 6.8\,arcsec.  
The ellipses mark the positions, sizes, and orientations of
discussed radio sources (taken from HyperLeda database). 
}
\label{tp6u}
\end{figure}

\begin{figure}
\resizebox{\hsize}{!}{\includegraphics{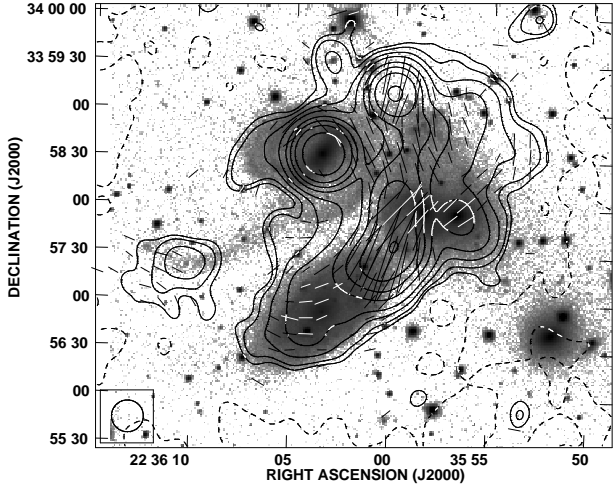}}
\caption{
Briggs weighted VLA D-array map of the total power emission of Stephan's
Quintet at 4.86\,GHz with apparent B--vectors of the polarised
intensity overlaid upon an SDSS-R image. The contour
levels are $-3$ (dashed), $3,5,10,20,50,100,200,500\times 6\,\mu \mathrm{Jy/beam}$,
(r.m.s. noise level). A B--vector of 1\,arcsec length
corresponds to a polarised intensity of $1.5\,\mu \mathrm{Jy/beam}$. 
The clip limit for the vectors is $15\,\mu \mathrm{Jy}$ (2.5 $\times$ PI noise level).
The angular resolution of the map is 20\,arcsec.}
\label{tp6n}
\end{figure}

The total power (TP) maps of HCG\,92 at 4.86\,GHz are shown in Fig.~\ref{tp6u} and
Fig.~\ref{tp6n}. The high resolution map (HPBW of 6.8\,arcsec)
shows two bright point sources and some extended emission. 

The most important structure in this study is the intergalactic emission
ridge located between the galaxies forming the group,
near $\mbox{R.A.}_{2000} = \mathrm{22^h 36^m 00^s}$ and  $\mbox{Dec}_{2000} = +33^{\circ} 57' 30''$.
In the observations by \citet{xuold}, the structure of the ridge is similar to one presented in
Fig.~\ref{tp6u}. Integrating the flux within the same boundaries,
we have obtained $10.6 \pm 0.6$\,mJy, very similar to their value
of $10.9 \pm 1.1$\,mJy.

The brightest source in the map (located at $\mathrm{R.A._{2000}}=22^\mathrm{h}
36^\mathrm{m} 04^\mathrm{s}$, $\mathrm{Dec_{2000}}=+33^{\circ} 58' 33''$)
is the core of the Seyfert type 2 galaxy NGC\,7319. The core is barely resolved 
and its flux of $9.38 \pm 0.49$\,mJy agrees with the value given by \citet{aoki}.

The $0.61~\pm~0.03$\,mJy peak at $\mathrm{R.A._{2000}}=22^\mathrm{h} 35^\mathrm{m} 57^\mathrm{s}$,
$\mathrm{Dec_{2000}}=+33^{\circ} 57' 55''$ is the core of NGC\,7318A. The flux is higher than presented
by Xu ($0.44 \pm 0.03$\,mJy) which may indicate the presence of an extended structure or variability. 

In the northern
part of the group, a strong point source near
$\mathrm{R.A._{2000}} = 22^\mathrm{h} 36^\mathrm{m} 00^\mathrm{s}$, 
$\mathrm{Dec_{2000}} = +33^{\circ} 59' 12''$
can be seen. The peak represents an unresolved double radio source,
denoted SQ-R by \citet{xuold}, most probably unrelated to
the group. Our 4.86\,GHz total flux of $4.0 \pm 0.2$\,mJy remains in 
very good agreement with the value of $3.7 \pm 0.4$\,mJy obtained 
by \citet{xuold}. 

South of SQ-R, a weak peak near $\mathrm{R.A._{2000}} = 22^\mathrm{h} 
35^\mathrm{m} 56^\mathrm{s}$ and
$\mathrm{Dec_{2000}} = +33^{\circ} 59\arcmin 20\arcsec$ represents the SQ-A
starburst region. Its flux of $0.32 \pm 0.02$\,mJy is in excellent
agreement with $0.3 \pm 0.1$\,mJy obtained by \citet{xuold}.

More extended emission can be seen in the lower resolution map (Fig.~\ref{tp6n}).
The second starburst region, SQ-B (which is likely a part of the tidal arm -- a remnant
of the past interactions with NGC\,7320C), located at  $\mathrm{R.A._{2000}} 
=22^\mathrm{h} 36^\mathrm{m} 10^\mathrm{s}$, $\mathrm{Dec_{2000}}=33^{\circ}57
\arcmin 22\arcsec$ has an integrated flux of $0.16 \pm 0.01$\,mJy, which is similar to
the value given by Xu ($0.2 \pm 0.1$\,mJy). Comparison with our uniformly
weighted map (Fig.~\ref{tp6u}) shows the presence of the extended, diffuse emission in that
region.

In the southern part of the group, the diffuse emission terminates near
the outskirts of the interloper galaxy NGC\,7320. The {\rm H}{\sc i} 
emission studies
by \citet{hi} show that the southern part of the group is connected
to an {\rm H}{\sc i} 
tail, extending eastwards from the group and containing the SQ-B
starburst region. Unfortunately, our map does not allow a reliable discrimination
whether or how the radio emission is connected to the
interloper galaxy (this issue is discussed in Sect.~\ref{tail1}).

The envelope of diffuse emission has an extension towards the western edge, which has 
no counterpart in the optical regime. The extension
is located near $\mathrm{R.A._{2000}} = 22^\mathrm{h} 35^\mathrm{m} 53^\mathrm{s}$, $\mathrm{Dec_{2000}} = 
33^{\circ}  58\arcmin 30\arcsec$. The lower resolution map shows
another extension -- towards the eastern part of the group,
overlapping the spiral arm of NGC\,7319.

\subsection{Distribution of the polarised intensity at 4.86\,GHz}
\label{sec6pi}
\begin{figure}
\resizebox{\hsize}{!}{\includegraphics{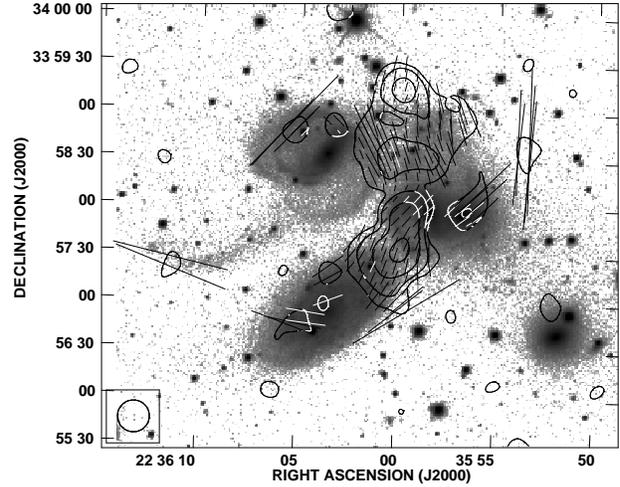}}
\caption{
Briggs weighted contour map of the polarised intensity distribution of
Stephan's Quintet at 4.86\,GHz with apparent B--vectors of the polarised
intensity overlaid upon an SDSS-R image.
The contour levels
are $-3$ (dashed), $3,5,10,20\times 6\,\mu \mathrm{Jy/beam}$ (r.m.s. noise level). A
polarisation vector of 1\,arcsec  corresponds to a 
polarisation degree of 0.5 per cent. 
The angular resolution of the map is 20\,arcsec.
}
\label{pi6}
\end{figure}

Figure~\ref{pi6} shows contours of the polarised intensity (PI) distribution
overlaid on a
greyscale optical image with the apparent B--vectors proportional to
the polarisation degree. The most prominent structure is the
emission ridge with a mean polarisation degree around 4--5 per cent.
The ridge is a part of an extended structure, filling also a large 
volume between NGC\,7318A and NGC\,7319 as well as the SQ-R source.
The polarisation degree of the latter is approximately 3 per cent.

Whereas the core of NGC\,7319 and the galaxy itself seem to be unpolarised,
the core of NGC\,7318A is present in the PI distribution map, with a polarisation
degree of approximately 6 per cent (averaged over the central area). 
The 4.86\,GHz PI distribution shows also a large pool of polarised
emission north from the ridge. The western extension of the TP envelope is 
also visible in our PI map.
 
East of the group, a spot of polarised emission spatially coincident
with SQ-B can be seen. The polarisation fraction of this source reaches 33 per cent.

\begin{figure}
\resizebox{\hsize}{!}{\includegraphics{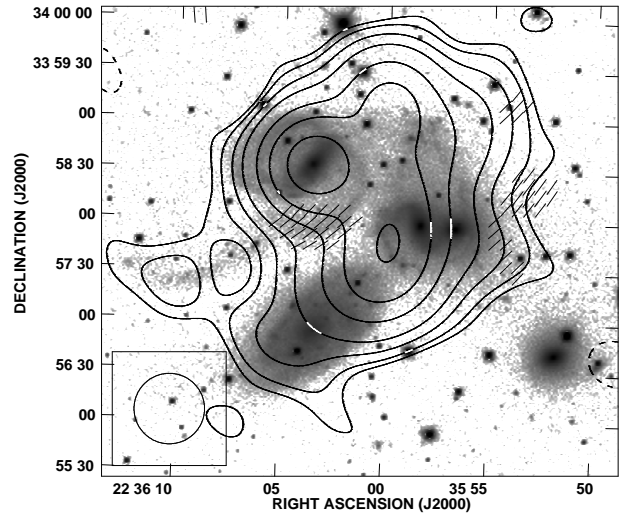}}
\caption{
Contour map of the total power emission of Stephan's Quintet at 1.43\,GHz 
with apparent B--vectors of the polarised intensity overlaid upon an SDSS-R image. 
The contour levels are $-3$ (dashed), $3,5,10,20,50,100,200\times 110\,\mu \mathrm{Jy/beam}$ 
(r.m.s. noise level). A polarisation vector of 1\,arcsec corresponds to a polarised intensity of
$10\,\mu \mathrm{Jy/beam}$.
The clip limit for the vectors is $80\,\mu \mathrm{Jy}$ (2.5 $\times$ PI noise level).
The angular resolution of the map is 42\,arcsec.    
}
\label{tp20}
\end{figure}

\subsection{Total power and polarised emission at 1.43\,GHz}
\label{sec20cm}
The TP map of HCG 92 at 1.43\,GHz with superimposed apparent polarisation B--vectors
is shown in Fig.~\ref{tp20}. The resolution
is considerably lower and this enables to trace the emission further out. 
Two point sources can be easily
distinguished from the surrounding emission, namely the core of NGC\,7319 and SQ-R.
The flux of the first one, estimated by a Gaussian fit centred on the galactic core, 
is $32 \pm 2$\,mJy and is therefore slightly higher
than 28.5\,mJy given by \citet{aoki} and by \citet{xuold}. 

Polarised emission from the ridge has not been detected at this frequency. Instead, weak emission
is present in the tidal tail of NGC\,7319. However, due to large
depolarisation (Sect.~\ref{depol}), we were unable to produce 
sufficiently reliable maps of the polarised intensity. 

\subsection{Total power and polarised emission at 8.35\,GHz}
\label{sec3cm}

\begin{figure}
\resizebox{\hsize}{!}{\includegraphics{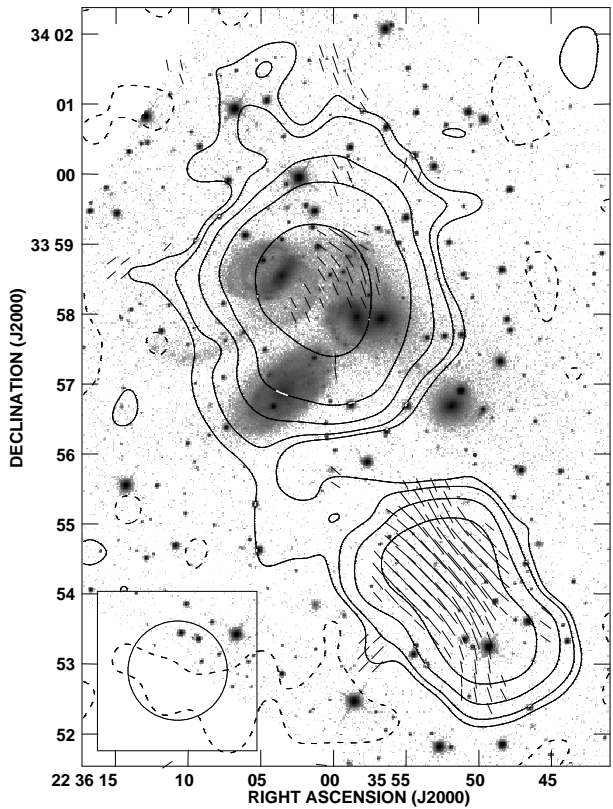}}
\caption{
Contour map of the TP emission of Stephan's Quintet at 8.35\,GHz,
with apparent B--vectors of the polarised intensity overlaid upon an SDSS-R image.
The contour levels are
$-3$ (dashed), $3,5,10,20,50 \times 100\,\mu \mathrm{Jy/beam}$ (r.m.s. noise
level). A polarisation
vector of 1\,arcsec corresponds to a polarised intensity 
of $20\,\mu \mathrm{Jy/beam}$.
The clip limit for the vectors is $180\,\mu \mathrm{Jy}$ (2.5 $\times$ PI noise level).
The angular resolution of the map is 85\,arcsec.  
}
\label{tp3}
\end{figure}

The TP emission distribution at 8.35\,GHz is shown in Fig.~\ref{tp3}.
Despite the lower (compared to the VLA) resolution,
one can easily notice that the total emission contours correspond fairly well to 
the ones seen at 1.43 and 
4.86\,GHz. The polarised emission exceeds the 3\,$\sigma$ 
r.m.s. level only at distinct regions in the group area, like between NGC\,7319
and NGC\,7318A.

The strong and polarised background source J223552+335425 located south of
Stephan Quintet was used to determine the foreground rotation measure (RM).
The best $\lambda^2$ fit to the polarisation angles of this source at all three
frequencies (1.43, 4.86, and 8.35\,GHz) yields a RM of 182\,rad\,m$^{-2}$, 
which agrees well with the values measured in the vicinity of the group \citep*{rm}.
With a value of 182\,rad\,m$^{-2}$ the polarisation angle
is rotated more than 360\degr{} at 1.43, 40\degr{} at 4.86,
and not more than 14\degr{} at 8.35\,GHz.

\section{Discussion} 
\label{discus}
\subsection{Spectral index}
\label{spixtext}
\begin{figure*}
\includegraphics[width=.5\textwidth]{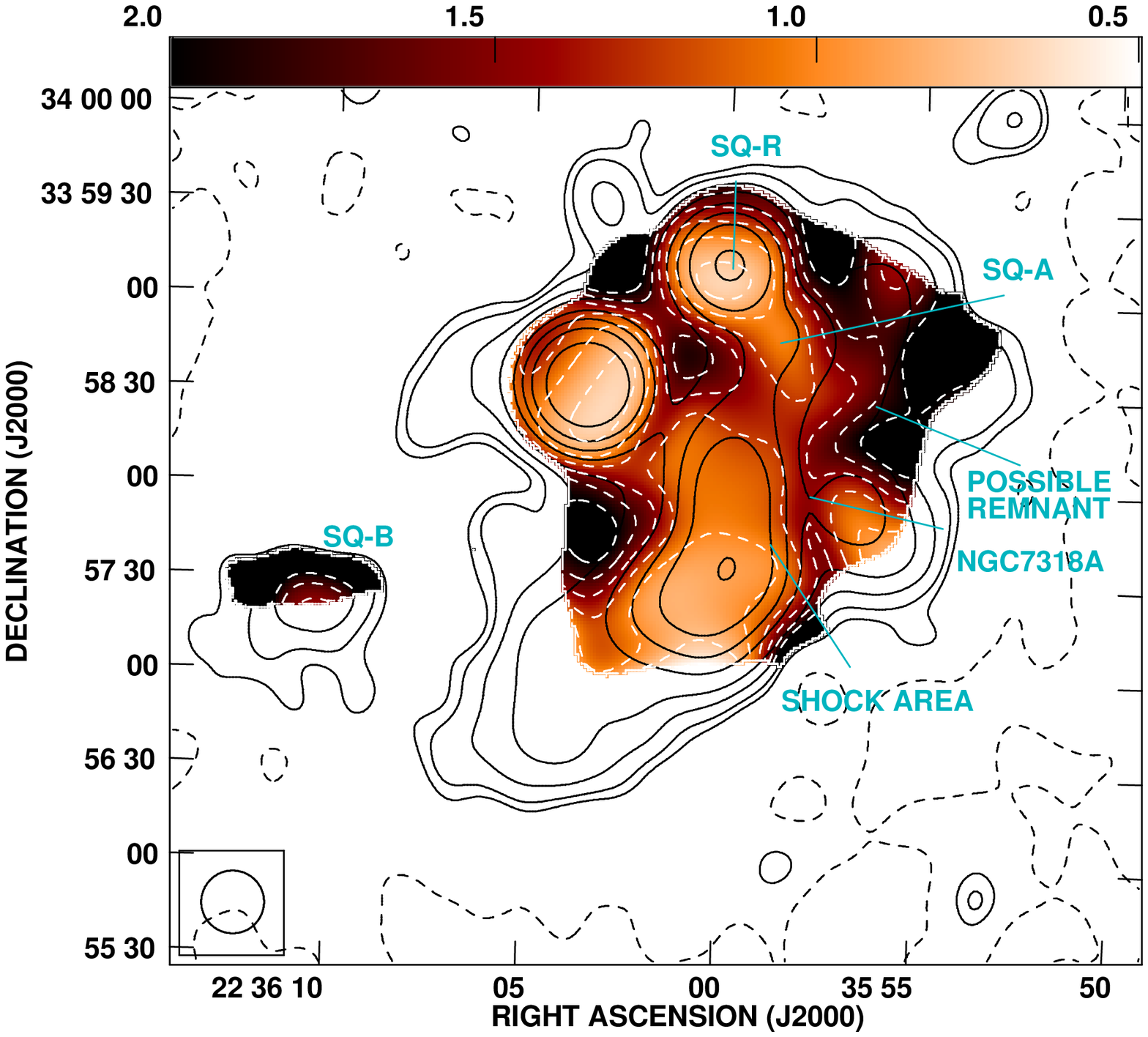}\hfill
\includegraphics[width=.5\textwidth]{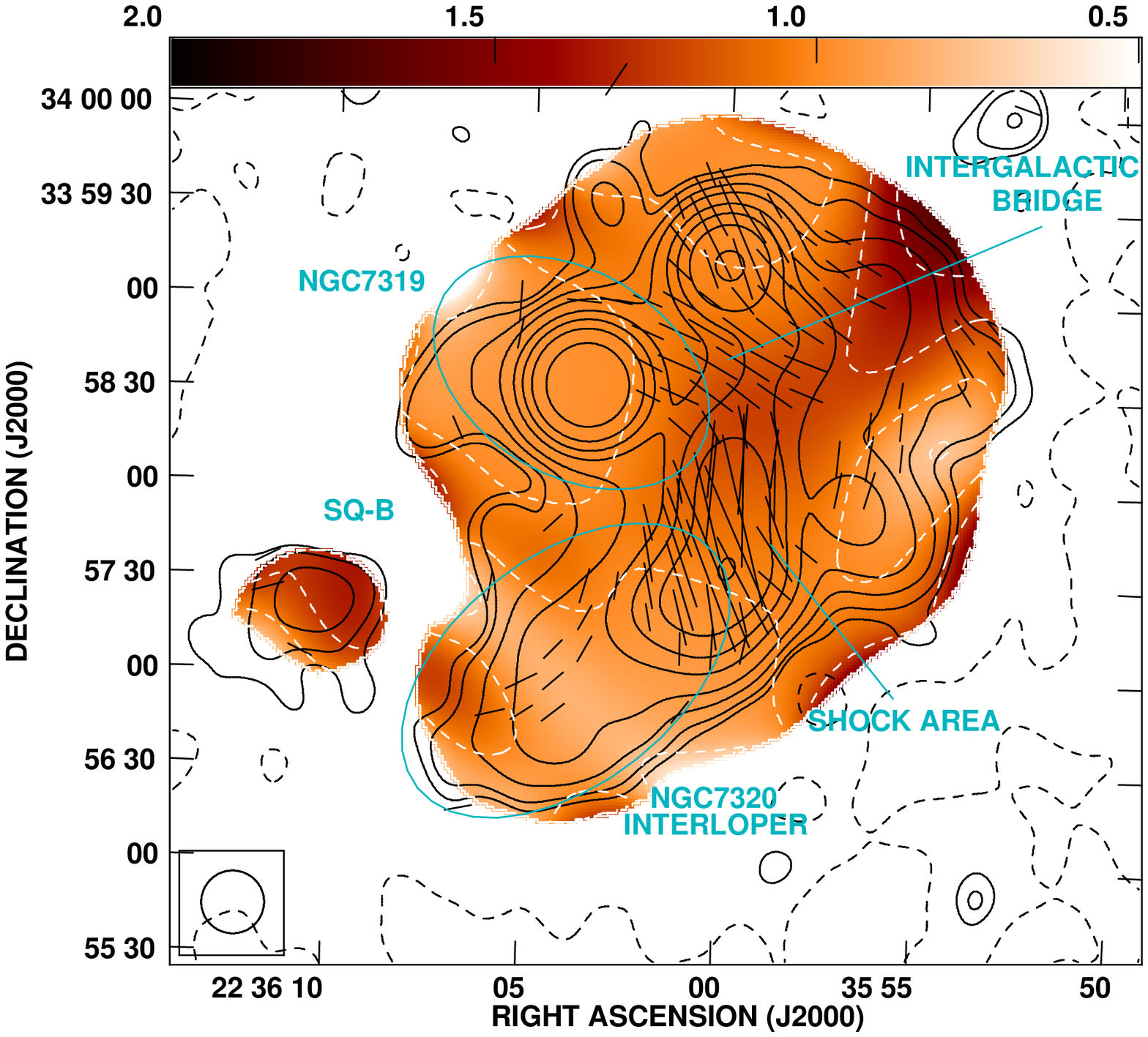}
\caption{
Distribution of the
spectral index in Stephan's Quintet calculated from 4.86 and 1.43\,GHz data
superimposed with black contours for the total power emission at 4.86\,GHz and (dashed) white contours for the spectral
indices.\newline
{\bf Left panel:} High resolution (20\,arcsec) map; {\bf Right panel:} Low resolution (42\,arcsec) map with B--vectors of the magnetic field
at 4.86\,GHz corrected for the foreground Faraday rotation. 
{\bf In both panels} the radio emission contour levels are $-3,3,5,10,20,50,100,200,500 \times 6\,\mu \mathrm{Jy/beam}$
(r.m.s. noise level). The spectral index contour levels are $2.0, 1.75, 1.5, 1.25, 1.0, 0.5$.
A polarisation vector (in the right panel) of 1\,arcsec corresponds to a polarised intensity of
$4.5\, \mu \mathrm{Jy/beam}$.
The clip limit for the vectors is $15\,\mu \mathrm{Jy}$ (2.5 $\times$ PI noise level).
}
\label{spixmap}
\end{figure*}

In order to calculate the spectral index distribution, we added to our 1.43\,GHz 
observations the datasets obtained by \citet{xuold} in VLA B-configuration at the same frequency,
taken from the NRAO archive. Combination of all available data at 1.43\,GHz yields maps with similar
resolution as at 4.86\,GHz. Maps at both frequencies were convolved with a
Gaussian function to obtain a final circular beam with a HPBW of 20\,arcsec
(small enough to distinguish point-source emission
from the extended one).
An additional map, with a beam of 42\,arcsec, has been made in
order to provide information about the regions of weak, diffuse emission,
not visible in the high resolution map. The spectral index
distribution is shown in Fig.~\ref{spixmap}.
Throughout the paper we are using 
the $S_{\nu} \propto \nu^{-\alpha}$
definition of the spectral index $\alpha$.

\begin{table}
\caption{\label{spixtbl} Spectral indices calculated for the Quintet}
\begin{center}
\begin{tabular}{lrrl}
\hline
\hline
Emission & $\alpha_{1.43-4.86\,GHz}$&$\alpha_{4.86-8.35\,GHz}$ \\
\hline
Total & 1.1 $\pm$ 0.2&1.1 $\pm$ 0.2\\
Point & 0.8 $\pm$ 0.2&0.6 $\pm$ 0.2\\
Diffuse & 1.2 $\pm$ 0.2&1.7 $\pm$ 0.2\\
\hline
\end{tabular}
\end{center}
\end{table}

\subsubsection{Point sources}
\label{spixpnt}
The spectral index of the core of NGC\,7319 --  approximately $0.83 \pm 0.07$ --
is, within the errors, consistent with the value given by \citet{xuold}. The 
lower resolution map allows to compute the spectral index also in the area of the 
spiral arm, where $\alpha=0.95 \pm 0.04$, so a value reasonable for galactic synchrotron emission.

The resolution of 20\,arcsec is high enough to measure the spectral index
for the background source SQ-R. The value of $0.81 \pm 0.09$ agrees very well
with $0.85 \pm 0.12$ given by \citet{xuold}.

The core of NGC\,7318A is surrounded by steep-spectrum emission.
The spectral index derived for the region close to the peak of the emission is $1.1 \pm 0.1$,
being substantially higher than the value given by Xu ($0.62 \pm 0.07$),
possibly indicating the presence of a diffuse, steep-spectrum component in our 1.43\,GHz map.

\subsubsection{Star Formation Regions}
\label{spixregs}
SQ-B region is marginally visible in the high resolution map,
but it can be clearly seen in the lower resolution image. The spectral
index of this region is approximately $1.2 \pm 0.2$; a value
somewhat higher than measured by \citet{xuold} -- $0.7 \pm 0.4$ -- but
still within the measurement uncertainties. Our low resolution TP map shows
more extended diffuse emission. As the beam is about three times larger than
in the study cited, the integrated flux value represents now not only the 
point source, but has also a contribution from the diffuse structure around. 
A diffuse, steep-spectrum emission ($\alpha$ = 1.4) contribution of
about 50 per cent in our larger beam map would explain the difference.

The area of the flatter spectrum within the intergalactic emission,
which is spatially correlated with the SQ-A region, can be distinguished from the 
surrounding emission only in the
high resolution map. The value obtained for the region ($1.1 \pm 0.1$)
is somewhat higher than given by Xu ($0.8 \pm 0.3$), but as in case of SQ--B,
within the given uncertainties. This again may be due to a contamination of 
extended, steep spectrum emission, like in the case of SQ-B.

\subsubsection{Intra-group emission}
\label{spixrid}
The emission ridge is clearly visible both in high and low resolution maps.
The spectrum steepens from approximately $0.85$ near
the southern boundaries to $1.1$ -- $1.3$ in the central part.
The mean spectral index of this region is equal to $1.1 \pm 0.15$ and
is consistent within the errors to the value obtained by Xu ($0.93 \pm 0.13$).

The 20\,arcsec resolution is sufficient to
separate the point sources from the diffuse emission, allowing
to determine the spectral index. The values vary from $1.3$
in the central part to $1.8$ in the outskirts, and are typical for an ageing population 
of synchrotron electrons.

The western extension coincides with a region  that shows a significantly
steeper spectrum with a mean index of $2.0 \pm 0.2$. 
Such steep spectrum indicates that this region can be a remnant of a past interaction 
within the group, as it shows effects of spectral ageing. 
The PI distribution map (Fig.~\ref{pi6}) shows a high degree of
polarisation of that region, approximately 40 per cent.

The mean spectral index of the whole diffuse emission of the group has been 
determined by calculating the total flux of the radio envelope in each of the three 
frequencies (taking into account only those
regions that are visible over the 3\,$\sigma$ r.m.s. level) and then subtracting
the flux of all point sources (taken from \citealt{aoki} and \citealt{xuold}). 
The obtained  values are given in Table \ref{spixtbl}. 
The table shows a steepening of the diffuse emission spectrum with increasing frequency. This is a
phenomenon typical for ageing particles in the intergalactic medium (as mentioned above,
the steepening occurs due to electron energy losses, see e.g. \citealt{pachol}). 
Any significant thermal component would manifest a substantial flattening of the spectrum between
higher frequencies, which is not the case here.

\subsubsection{{\rm H}{\sc i} tail or an interloper galaxy?}
\label{tail1}
The low resolution map (Fig.~\ref{spixmap}, right panel) was made
to visualise the spectral index of the diffuse
emission regions, which have a surface brightness too low to be visible in
the high resolution map. One of them is the southern extension,
where the H{\sc i} tail overlaps the galaxy NGC\,7320 \citep{hi}.
The measured $\alpha$ of approximately $0.9\,\pm\,0.1$ 
is a relatively flat spectrum (similar to NGC\,7318A and NGC\,7319),
suggesting that the emission is related rather to the star--forming interloper galaxy than 
to an intergalactic structure. The polarised
fraction of this region reaches 13 per cent, as mentioned in Sect.~\ref{sec6pi}.
Fig.~\ref{nuv_vec} demonstrates the alignment of the vectors in
the southern tail of the group. The vectors form an arc bending in
the eastern direction, matching the {\rm H}{\sc i} tail. The arc itself is
connected to the vectors overlapping the shock area. However, as
the inclination of NGC\,7320 is approximately $60\degr$, this may
only be a projection of the magnetic field lines of the foreground galaxy.

NGC\,7320 is an example of a late-type, slowly rotating, low-mass system (\citealt{hyperleda}),
similar to the galaxies described by \citet{chyzy}. Such galaxies are characterised by a relatively flat
spectral index ($\alpha \simeq 0.5$). If such radiation is overlapping the one
from the {\rm H}{\sc i} tail, a combination of the two components (assuming that 30\,\%
of the flux comes from interloper galaxy) would explain the observed value of the spectral index.

\subsection{Faraday Depolarisation}
\label{depol}

As the PI distribution map at 1.43\,GHz shows very weak 
polarised emission (see Sect.~\ref{sec20cm}), we decided to estimate the
depolarisation level between 1.43 and 4.86\,GHz. 
We use the term ``depolarisation'' defined as $DP = 1 - {p}_{1.43}/{p}_{4.86}$, where
$p$ refers to the polarisation degree at a given frequency.
As both maps were made with the same beamsize, such a definition makes this parameter independent of beam depolarisation.
The mean depolarisation of the radio--emitting envelope between 1.43 and 4.86\,GHz is equal to 0.84.
In contrast, the background source J223552+335425 shows depolarisation
of only 40 per cent.

The exact values can be affected by differences in the bandwidth depolarisation
at both frequencies, but this effect 
cannot affect the variation of $DP$ with position in the map.
These values indicate that the depolarisation caused by the foreground 
Faraday dispersion is not likely to explain the high degradation of the polarised emission
estimated for the Quintet. The alternative explanation is
that the depolarisation in the group is caused either by Faraday rotation
\textit{inside} Stephan's Quintet or by internal Faraday dispersion.
The first possibility is that the intergalactic space in the Quintet hosts a
substantial unidirectional, dynamo--type magnetic field. However, the present data
do not allow to disregard the scenario of depolarisation via Faraday dispersion.
Moreover, there is a possibility that the depolarisation estimate for the background source 
is influenced by its internal depolarisation. Furthermore, the foreground
depolarisation distribution might be patchy.
The RM Synthesis method (as described by \citealt{rmsynth}) will
probably be able to distinguish between those two mechanisms, but it needs much 
better frequency coverage and higher resolution than offered by the existing data. 

\subsection{Magnetic field strengths in the Quintet}
\label{strenght}

The strength of the magnetic field and its energy density were calculated from the 4.86\,GHz
data, assuming energy equipartition between the cosmic rays and the
magnetic field, following the formulae presented in \citet {bfeld}.
The chosen values of the parameters (total pathlenght $D$, proton--to--electron
energy density ratio $K_0$, spectral index $\alpha$, 4.86 \,GHz flux $S_{4.86}$, and the polarisation
degree $p$) for each of the regions are presented in Table \ref{params}.
The same table contains also calculated total and ordered field strength as well
as the magnetic field energy density in each case.

\begin{table*}
\caption{Parameters used to estimate the magnetic field properties and resulting values}
\begin{center}
\begin{tabular}[]{lcccccccc}
\hline
\hline
Region & {\it D}[kpc]& $K_0$ & $\alpha$ & $S_{4.86}$\,[mJy]& $p$\,[$\%$]
 &$B_{\mbox{TOT}}$ [$\mu$G]&$B_{\mbox{ORD}}$ [$\mu$G]&$E_{\mbox{B}}$ [$\mathrm{erg}\,\mathrm{cm^{-3}}$] \\
\hline
Ridge & 12.5 $\pm$ 2.5 & 100 $\pm$ 50 & 1.1 $\pm$ 0.15 & 10.6 $\pm$ 0.6 & 5
& $11.0 \pm 2.2$&$2.6 \pm 0.8$&$0.5 \pm 0.15 \times 10^{-11}$\\
SQ-A & 6 $\pm$ 3 & 100 & 1.1 $\pm$ 0.1 & 0.32 $\pm$ 0.02 & ------
& $8.8 \pm 2.3$&------------&$3.0 \pm 1.6 \times 10^{-12}$\\
SQ-B & 6 $\pm$ 3 & 100 & 1.2 $\pm$ 0.2 & 0.16 $\pm$ 0.01 & 33
& $6.5 \pm 1.9$&$3.5 \pm 1.2$& $1.8 \pm 0.9 \times 10^{-12}$ \\
Group & 32 $\pm$ 6 & 100 & 1.2 $\pm$ 0.2 & 4.6 $\pm$ 0.6 & 2
&$6.4 \pm 1.1$&$1.1 \pm 0.3$&$1.8 \pm 0.5 \times 10^{-12}$ \\
\hline
\end{tabular}
\end{center}
\label{params}
\end{table*}

\subsubsection{Magnetic field in the ridge}

The pathlength $D$ through the ridge was estimated
to be 10--15\,kpc (based on the size of the shock region from the high resolution radio map
and assuming cylindrical symmetry). The flux and spectral index were taken from our VLA data. As
the spectrum is relatively steep, the thermal fraction contribution is negligible.
The problem arises with the value of the $K_{0}$ coefficient. Its value depends 
on the strength of a shock and for strong shocks (compression ratio $r \geq 3.4$) 
$K_0$ reaches 40--100.
Assuming $K_{0} = 100 \pm 50$, the total magnetic field strength in the ridge is 
$11.0\,\mu\mbox{G} \times \left(\frac{K_0}{100}\right)^{0.244}\pm 2.2\,\mu$G 
with an ordered component of $2.6 \pm 0.8\,\mu$G. It is, however, not certain if the shock in the Quintet is a 
strong or a weak one, as weak shocks would also be able to produce the observed X-ray properties
of the group (see \citealt{osull}, Sect.~5.3, for a detailed discussion). In that case, the $K_{0}$ value would be 
much higher (even by several orders of magnitude). For relatively weak shocks ($r \leq 2.2$), the 
magnetic field strength would increase more than three times.

The total magnetic energy density of the shock area was estimated as $E_{\mathrm{B}} = 0.5 \pm 0.15 \times 10^{-11}
\mathrm{erg}\,\mathrm{cm^{-3}}$. The thermal energy of the shock area was estimated from
the X-ray data using the temperature of 0.6\,keV and gas density of $1.167 \times 10^{-2}$ cm$^{-3}$ taken from
\citet{osull}, yielding value of $\approx 1.1 \times 10^{-11} \mathrm{erg}\,\mathrm{cm^{-3}}$.
This means that the magnetic field plays an important role in the dynamics of the intergalactic medium (IGM) -- contrary to the
statement by \citet{xuold}.
Its contribution to the total energy, comparable to the thermal component, proves that 
it is necessary to take the magnetic field into
account while performing simulations of the intra-group medium.

\subsubsection{Magnetic field in the tidal dwarf galaxy}

Stephan's Quintet is known to be a host group for at least 13
TDG candidates, located in the tidal tail connected to NGC\,7319
\citep{bravedwarfs}. One of them is the star 
formation region SQ-B.
For SQ-B we calculated the properties of the magnetic field
under the assumption that the $K_0$ ratio is equal 
to 100, as we  do not expect the field to be produced by secondary
electrons or shocks (see \citealt{bfeld} for details).
The total pathlength through the emitting volume
was chosen to be equal to $6 \pm 3$\,kpc. The flux and spectral index 
were taken from our VLA data. As the spectrum is relatively steep ($1.2 \pm 0.2$),
the thermal fraction contribution is negligible.

The magnetic field of the TDG candidate is relatively strong ($\approx 6.5\,\mu$G),
similar to the ones found in normal-sized spiral galaxies, for which the median
value is $9 \pm 1.3\,\mu$G \citep{niklas}. 
The ordered component (reaching $3.5 \pm 1.2\,\mu$G) is also significant,
as polarisation degree is substantial (33 per cent).

The strength of the magnetic field and its anisotropy suggest an
\textit{in situ} amplification of the field in the TDG candidate.
The presence of the star formation and the shearing flow of re-accreting
of plasma debris left by the passage of NGC\,7320C can efficiently
amplify the magnetic field even if we assume a low mass and slow rotation of a dwarf galaxy
\citep{hubnew}. Given that the debris plasma is likely 
to have already been magnetised, the amplification process could 
result in values as high as in normal galaxies.

The second starburst region, SQ-A, lies inside an extended polarised region.
Hence, it cannot be distinguished from the surrounding emission.
The total magnetic field strength 
of this region was estimated as $8.8 \pm 2.3\,\mu$G. This translates into a magnetic field energy density 
of $3.0 \pm 1.6 \times 10^{-12} \mathrm{erg}\,\mathrm{cm^{-3}}$.

\subsubsection{Mean magnetic field in the group area}

In order to estimate the mean magnetic field in the group area, we have subtracted 
the point sources (from \citealt{aoki}) and the ridge area,
and then clipped the resulting map at the level of approximately $8\,
\sigma$ to obtain the integrated flux of the group. The pathlength was adopted as 
equal to the separation 
between the galaxies originally forming the Quintet (NGC\,7317, NGC\,7318A, and NGC\,7319), 
$K_0$ was adopted as 100. The spectral
index was taken from our VLA data. As the spectrum is relatively steep,
we decided to neglect the thermal flux contribution. We estimated 
the strength of total magnetic field 
to be $6.4 \pm 1.1\,\mu$G with an ordered component of $1.1 \pm 0.3\,\mu$G, 
indicating an energy density
of $1.8 \pm 0.5 \times 10^{-12} \mathrm{erg}\,\mathrm{cm^{-3}}$.
The magnetic field in the intergalactic space is of similar strength as in 
the star forming regions SQ-A and SQ-B.

\subsubsection{Shock compression as a possible origin of the magnetic field
in the ridge}

\begin{figure}
\resizebox{\hsize}{!}{\includegraphics{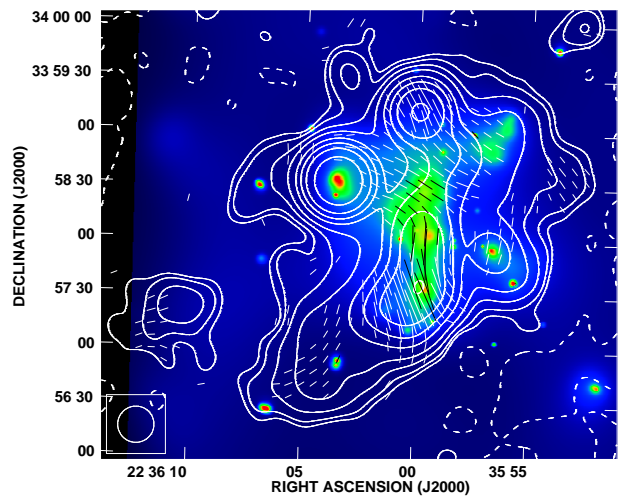}}
\caption{
Map of B--vectors of the polarised intensity at 4.86\,GHz after the
correction for the foreground Faraday rotation overlaid upon the soft X-ray image
from \textit{CHANDRA}. The contour levels are
$-3$ (dashed), $3,5,10,20,50,100,200,500\times 6\,\mu \mathrm{Jy/beam}$ (r.m.s. noise level).
A polarisation
vector of 1\,arcsec corresponds to a polarised intensity of
$4.5\, \mu \mathrm{Jy/beam}$.
The clip limit for the vectors is $15\,\mu \mathrm{Jy}$ (2.5 $\times$ PI noise level).
The angular resolution of the radio data is 20\,arcsec. 
}
\label{chnd_cont}
\end{figure}

\begin{figure}
\resizebox{\hsize}{!}{\includegraphics{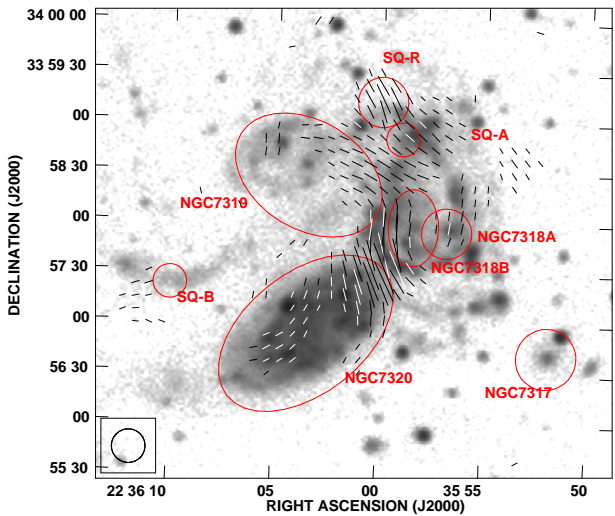}}
\caption{
Map of B--vectors of the polarised intensity at 4.86\,GHz after the
correction for the foreground Faraday rotation overlaid upon the near UV image from \textit{GALEX}.
A polarisation vector of 1\,arcsec corresponds to a polarised intensity of
$4.5\,\mu \mathrm{Jy/beam}$.
The clip limit for the vectors is $15\,\mu \mathrm{Jy}$ (2.5 $\times$ PI noise level).
The angular resolution of the radio data is 20\,arcsec. 
The ellipses mark the positions, sizes and orientations of
discussed radio sources (taken from HyperLeda database).
NGC\,7320C galaxy lies approximately $1\degr$ away from the eastern boundary of this image.
}
\label{nuv_vec}
\end{figure}

The emission ridge has been studied extensively in different regimes
of the electromagnetic spectrum (see references in Sect.~\ref{intro}).
Three mechanisms explaining the observed properties of the ridge
have been described and then
tested for the Quintet. Proposed were the accretion of the primordial gas
(\citealt{osmond}, \citealt{osull}), heating of the medium by high-mass
X-ray binaries and Supernovae \citep{osull}, and shock heating (\citealt{jabko},
\citealt{osull}). The last of them was suggested as the most probable one,
providing the most accurate explanation of the observed energy and 
temperature distributions of the X-ray emitting medium, as well as explaining
phenomena seen in other regimes of the electromagnetic spectrum.
The study of the magnetic field can provide arguments for or against
the shock scenario, as propagation of the shock waves through the
magnetised plasma should result in changing the orientation of the
magnetic field; in particular, the magnetic field can thus be squeezed,
resulting in higher polarisation degree \citep{urb}. 
As the emission ridge is expected to be formed by
the means of shocking the IGM due to high-speed infall of NGC\,7318B 
to the group, there should be a polarised structure between the galaxies
mentioned above. In Fig.~\ref{chnd_cont}, the TP emission contours
and B--vectors of the polarised intensity, corrected for the foreground Faraday rotation,
from our 4.86\,GHz observations have been overlaid upon the \textit{CHANDRA} image
presenting the X-ray emission. The area of the shock agrees well
with the polarised radio ridge, with the maximum located near the southern
end of the X-ray ridge, indicating an increased polarisation degree of the IGM
in that region. As the shock is clearly visible in the UV data, we
have superimposed the \textit{GALEX} near UV image of the Quintet with the 
polarisation B--vectors (corrected for the foreground RM) at 4.86\,GHz (Fig.~\ref{nuv_vec}). 
The orientation of the vectors (parallel to the shock) 
strongly supports the idea of an enhancement of the anisotropy due to 
shock-driven compression.

\subsubsection{Magnetic field as a tracer of the previous interactions}
\label{7319}

Fig.~\ref{pi6} shows 
an extended area of enhanced polarisation degree between
the radio ridge and NGC\,7319 
(near $\mathrm{R.A._{2000}} = 22^\mathrm{h} 36^\mathrm{m} 00^\mathrm{s}$, $\mathrm{Dec_{2000}} = +33^{\circ} 58' 20''$).
The polarised fraction is significant, ranging 
from approximately 6 per cent near
the ridge up to 13 per cent in the outskirts of the western spiral arm of NGC\,7319.
The magnetic field, represented by the B--vectors 
(corrected for the foreground Faraday rotation of 182\,rad\,m$^{-2}$,
as seen in Fig.~\ref{nuv_vec}) seems to connect NGC\,7319
with the pair NGC\,7318A/B. The low resolution map of the spectral index (Fig.~\ref{spixmap}, right panel)
shows that $\alpha$ varies from approximately $1.2$ to $1.8$, with mean value of $1.45 \pm 0.15$.

NGC\,7319 is known to be perturbed by the previous interactions. It is usually 
suggested that they
were due to a hypothetical passage of NGC\,7320C 
(\citealt*{shostak}, \citealt*{molesold}, \citealt*{molesnew}) that
caused stripping of the material through the tidal tail containing SQ-B. 
On the other hand, NGC\,7318A
(as well as NGC\,7317) is considered as non-interacting member of the group.

The diffuse emission with a steep spectrum and a high degree of 
polarisation is likely to originate in the material stripped from NGC\,7319, as this galaxy
shows hardly any signs of the emission neither in {\rm H}{\sc i} (\citealt{shostak}, 
\citealt{hi}), H$\alpha$ (\citealt{arp}, \citealt{molesold}), nor in CO 
\citep{yun}. 
An active role of NGC\,7318A in the previous interactions was first proposed by \citet{shostak}
and later supported by \citet{xunew}, who proposed that the "UV Loop" structure connected to NGC\,7319 
might be a "counter-tidal" tail formed during an encounter.

\section{Conclusions \label{sec:conclusions}}

We observed Stephan's Quintet group of galaxies using the VLA at 1.43 and 4.86\,GHz and the
Effelsberg \mbox{100m} radio telescope at 4.85 and 8.35\,GHz. We obtained maps of total power emission and polarised intensity.
These maps were analysed together with the archive X-ray and UV data in order to explore 
the properties of the magnetic
field in the group. We conclude:
\begin{enumerate}
\item[--] The group has a large radio envelope, visible at 1.43, 4.86, and 8.35\,GHz. The envelope encompasses all the member galaxies.
\item[--] There is a narrow, S--shaped region of the radio emission between the member galaxies. It extends from
the background source at $\mathrm{R.A._{2000}} = 22^\mathrm{h} 36^\mathrm{m} 00^\mathrm{s}$, $\mathrm{Dec_{2000}} = +33^{\circ} 59' 11''$ towards the shock region and
diminishes near the northwestern edge of the foreground galaxy NGC\,7320. 
\item[--] The mean polarisation degree of the shock region is 5 per cent. The magnetic field strength obtained within this region is equal to $11.0\,\mu\mbox{G} \times \left(\frac{K_0}{100}\right)^{0.244} \pm 2.2\,\mu$G,
with an ordered component of $2.6 \pm 0.8\,\mu$G. The energy density of $0.5 \pm 0.15 \times 10^{-11} \mathrm{erg}\,\mathrm{cm^{-3}}$ is comparable
to the thermal one, indicating the dynamical importance of the magnetic field in the physics of the 
intra-group medium.
\item[--] The radio emission from the aforementioned envelope is polarised, with a mean polarisation degree of 2 per cent. The strength of the mean magnetic field 
within its boundaries is equal to $6.4 \pm 1.1\, \mu$G, with an ordered component of 
$1.1 \pm 0.3\, \mu$G. The average magnetic field energy density is
$1.8 \pm 0.5 \times 10^{-12} \mathrm{erg}\,\mathrm{cm^{-3}}$.
\item[--] The depolarisation
of the emission from the Quintet calculated from the 1.43 and 4.86\,GHz data
exceeds 80 per cent. This is more than two times higher than the depolarisation
of the neighbouring background source. Such a difference suggests depolarisation of the emission 
from the group either by intrinsic (within the emitting region) Faraday rotation
or internal Faraday dispersion. In the first case, it would indicate the presence of a
regular magnetic field.
\item[--] The intergalactic emission has a rather steep spectrum, with a mean spectral index of $1.2 \pm 0.2$ between 1.43 and 4.86\,GHz and 
$1.7 \pm 0.2$ between 4.86 and 8.35\,GHz. The steepness of the spectrum indicates that the intergalactic emission may be dominated by an 
ageing population of electrons and that the thermal component does not play a significant role.
\item[--] There is a region of a steep-spectrum ($2.0 \pm 0.2$), highly (40 per cent) polarised emission on the northwestern edge of the radio envelope. This region
might be a remnant of the past interactions among the group members.
\item[--] In the southern part of the group the emission forms an
extension overlapping the {\rm H}{\sc i} 
 tail (detected by \citealt{hi}).
Although the orientation of the B--vectors seems to follow the {\rm H}{\sc i} tail,
the spectral index of the emission ($0.9 \pm 0.1$) indicates that
it emerges not only from within the group, but also from the interloper galaxy NGC\,7320.
Moreover, the high inclination of NGC\,7320 may result in projecting its
magnetic field so that the B--vectors form an arc-like structure.
\item[--] The radio emission from the starburst region SQ-B is substantially
polarised (33 per cent), indicating presence of the magnetic
field with a total strength of $6.5 \pm 1.9\,\mu$G and an ordered
component reaching $3.5 \pm 1.2\,\mu$G. As this structure is supposed
to be an example of a tidal dwarf galaxy, the detected field is likely
intrinsic to the dwarf, amplified by the flow of re-accreting of stripped, magnetised
plasma from the neighbour galaxy NGC\,7319 during the passage of NGC\,7320C.
\end{enumerate}

\section*{acknowledgements}
We would like to thank an anonymous referee for helpful
comments and suggestions. We thank Kerstin Weis (RUB) and Marita Krause (MPIfR Bonn) for valuable comments.
This research has been supported
by the scientific grant from the National Science Centre (NCN), DEC. no.\,2011/03/B/ST9/01859.
DJB and RB acknowledge support by the DFG SFB\,591 `Universal 
Behaviour of non-equilibrium plasmas' and DFG FOR\,1254, `Magnetisation of Interstellar
and Intergalactic Media`.
This research has made us of
the NASA/IPAC Extragalactic Database (NED) which is operated by the Jet Propulsion Laboratory,
California Institute of Technology, under contract with the National Aeronautics and Space Administration. 
This research has made use of NASA's Astrophysics Data System.
Funding for the SDSS and SDSS-II has been provided by the Alfred P. Sloan Foundation, the Participating Institutions, 
the National Science Foundation, the U.S. Department of Energy, the National Aeronautics and Space Administration, 
the Japanese Monbukagakusho, the Max Planck Society, and the Higher Education Funding Council for England. 
The SDSS Web Site is http://www.sdss.org/.
This research has made use of data obtained from the Chandra Data Archive and software provided by the Chandra X-ray 
Center (CXC) in the application packages CIAO, ChIPS, and Sherpa. 
We acknowledge usage of the Galaxy EVolution EXplorer (GALEX).


\label{lastpage}

\end{document}